\documentstyle[emulateapj,psfig]{article}

\makeatletter

\newenvironment{inlinefigure}{%
\def\@captype{figure}%
\noindent\begin{minipage}{0.999\linewidth}\begin{center}}
{\end{center}\end{minipage}\smallskip}
\makeatother

\def\cle      {{$_ <\atop{^\sim}$}}

\newcommand{\mum}{$\,\mu$m}
\newcommand{\smm}{$S_{\rm 850\,\mu m}$}
\newcommand{\rad}{$S_{\rm 1.4\,GHz}$}
\newcommand{\sub}{submillimeter}
\newcommand{\ratio}{$S_{\rm 850\,\mu m}/S_{\rm 1.4\,GHz}$}

\lefthead{Chapman et al.}

\begin{document}

\title{The Properties of Microjansky Radio Sources in the
HDF-N, SSA13, and SSA22 Fields}
\author{S.~C.~Chapman,$\!$\altaffilmark{1}
A.~J.~Barger,$\!$\altaffilmark{2,3,4}
L.~L.~Cowie,$\!$\altaffilmark{4}\\
D.~Scott,$\!$\altaffilmark{5}
C.~Borys,$\!$\altaffilmark{5}
P.~Capak,$\!$\altaffilmark{4}
E.~B.~Fomalont,$\!$\altaffilmark{6}
G.~F.~Lewis,$\!$\altaffilmark{7}
E.~A.~Richards,$\!$\altaffilmark{8}
A.~T.~Steffen,$\!$\altaffilmark{2}
G.~Wilson,$\!$\altaffilmark{9}
M.~Yun$\!$\altaffilmark{10}
}

\altaffiltext{1}{California Institute of Technology, Pasadena, CA 91125}
\altaffiltext{2}{Department of Astronomy, University of Wisconsin-Madison, 
Madison, WI 53706}
\altaffiltext{3}{Department of Physics and Astronomy, University of Hawaii,
Honolulu, HI 96822}
\altaffiltext{4}{Institute for Astronomy, University of Hawaii,
Honolulu, Hawaii 96822}
\altaffiltext{5}{Department of Physics \& Astronomy,
University of British Columbia, Canada V6T\,1Z1}
\altaffiltext{6}{National Radio Astronomy Observatory, Charlottesville,
VA 22903}
\altaffiltext{7}{Anglo-Australian Observatory, P.O. Box 296, Epping, 
NSW 1710, Australia}
\altaffiltext{8}{University of Alabama, Huntsville, AL 35899}
\altaffiltext{9}{Physics Department, Brown University, Providence, 
RI 02912}
\altaffiltext{10}{University of Massachusetts, Department of Astronomy, 
Amherst, MA 01003}

\slugcomment{Submitted to The Astrophysical Journal} 

\begin{abstract}

We present multiwavelength observations for a large sample of
microjansky radio sources detected in ultradeep 1.4\,GHz maps
centered on the Hubble Deep Field-North (HDF-N) and the Hawaii 
Survey Fields SSA13 and SSA22. Our spectroscopic redshifts for 169
radio sources reveal a flat median redshift distribution,
and these sources are hosted by similarly luminous optical $L^\ast$
galaxies, regardless of redshift. We suggest that the absence
of low optical luminosity galaxies at low redshifts, where there
are no selection effects, is due to small galaxies
not being efficient at retaining the cosmic rays necessary to
host microjansky radio sources. 
This is a serious concern for radio
estimates of the local star formation rate density, as a substantial
fraction of the ultraviolet luminosity density is generated by
sub-$L^*$ galaxies at low redshifts.
From our submillimeter measurements for 278 
radio sources, we find error-weighted mean 850\mum\ fluxes 
of 1.72$\pm$0.09\,mJy for the total sample, 
2.37$\pm$0.13\,mJy for the optically-faint ($I>23.5$) subsample, 
and 1.04$\pm$0.13\,mJy for the optically-bright ($I<23.5$) subsample.
We significantly ($>3\sigma$) detect in the submillimeter 50
of the radio sources, 38 with $I>23.5$.
Spectroscopic redshifts for three of the 
$I<23.5$ submillimeter-detected radio sources are in the 
range $z=1.0$--3.4, and all show AGN signatures. 
Using only the submillimeter mapped regions 
we find that $69 \pm9$\% of the submillimeter-detected radio 
population are at $I>23.5$. We also find that $66 \pm7$\% of the
\smm$>5$\,mJy ($>4\sigma$) sources are radio-identified.
We use our spectroscopic sample to determine the evolution with
redshift of the radio power, and hence the far-infrared (FIR)
luminosity (through the local FIR-radio correlation). 
We find that millimetric redshift estimates at low redshifts are best 
made with a FIR template intermediate between a Milky Way type galaxy 
and a starburst galaxy, and at high redshifts with an Arp~220 template.
\end{abstract}


\section{Introduction}
\label{secintro}

Some of the deepest 1.4\,GHz maps ever obtained with the 
Very Large Array (VLA) are centered on the
Hubble Deep Field-North (HDF-N) and the Hawaii Survey 
Fields\footnote{Central coordinates: $13^{\rm h} 12^{\rm m} 35^{\rm s}.2$,
$42^\circ 44' 24''$ (SSA13) and $22^{\rm h} 17^{\rm m} 59^{\rm s}.2$,
$00^\circ 18' 31''$ (SSA22); Cowie et al.\ (1996).} SSA13 and SSA22.
These maps should be sensitive to the synchrotron
emitting disks of galaxies out to $z\sim1$ and the nuclear starbursts
of ultraluminous infrared galaxies 
(\markcite{sanders96}Sanders \& Mirabel 1996) out to $z\sim 3$.
Most importantly, since 1.4\,GHz emission does not suffer from 
absorption by galactic dust or neutral hydrogen at high redshifts, 
even the most distant, dust-obscured galaxies should be transparent. 
The 1.4\,GHz maps of the HDF-N, SSA13, and SSA22 fields
are not only unprecedented in their sensitivities
(e.g., the r.m.s.~of the SSA13 image is 5~$\mu$Jy) 
but also in their detailed 
resolution, which has enabled unambiguous optical identifications 
to be made on images aligned to within $0.1''-0.2''$ of the 
radio FK5 coordinate grid. These optical identifications are 
composed of 10\%$-20$\% late-type Seyfert galaxies and 
bright field elliptical galaxies and 60\%$-70$\% moderate redshift
($0.1<z<1.3$) starburst galaxies characterized by their diffuse
synchrotron emission on galactic scales ($1''-2''$;
T.~W.~B.~Muxlow et al., in preparation). 
The remainder of the sources are unidentified to $I\sim25$
Thus, the faint radio 
source population appears to be dominated by moderate-redshift 
star-forming galaxies 
(e.g., \markcite{windhorst95}Windhorst et al.~1995;
\markcite{richards00}Richards 2000).

Sensitive submillimeter follow-up observations with the 
Submillimetre Common User Bolometer Array (SCUBA;
\markcite{holland99}Holland et al.\ 1999) on the 15~m
James Clerk Maxwell Telescope\footnote{The
JCMT is operated by the Joint Astronomy Centre on behalf of the parent
organizations the Particle Physics and Astronomy Research Council in
the United Kingdom, the National Research Council of Canada, and
the Netherlands Organization for Scientific Research.}
of the optically-faint tail of the radio source population
have revealed significant numbers of \smm$>5$\,mJy sources
(Barger, Cowie, \& Richards 2000, hereafter BCR00;
\markcite{chapman01}Chapman et al.~2001, hereafter C01;
Chapman et al.~2002a, hereafter C02) whose submillimeter emission
appears to be produced primarily by star formation rather than
by accretion onto supermassive black holes
(\markcite{barger01a}Barger et al.\ 2001a, b).

A radio selection technique can therefore
be used to identify bright \sub\ sources without the need for 
long submillimeter exposures. C01 found that high-redshift 
\smm$>$5\,mJy sources can be detected
with SCUBA in photometry mode at a rate of about one per hour,
which is significantly more rapid than the rate of about one per
10~hours achieved in blank-field jiggle map surveys
(e.g., \markcite{barger98}Barger et al.\ 1998;
\markcite{hughes98}Hughes et al.\ 1998).
Moreover, the \ratio\ ratio can be used as a millimetric redshift 
indicator (\markcite{cy99}Carilli \& Yun 1999, 2000, hereafter CY99 
and CY00; BCR00; \markcite{dunne00}Dunne, Clements, \& Eales\ 2000, 
hereafter DCE00). Millimetric redshift estimates of
\sub-detected radio sources are typically in the range $1<z<3$ 
(e.g., BCR00; \markcite{smail00}Smail et al.\ 2000; C01; C02;
Ivison et al.\ 2002).

Flux-limited radio selection does introduce a luminosity bias
with redshift (see \S~\ref{radioev}); even very sensitive
radio maps, such as those used in the present paper,
will not detect galaxies at very high redshifts.
C02 modeled the radio selection function in detail using 
different galaxy evolution scenarios and found that
the primary bias was a cut-off of galaxies at $z$\cle 3 due to 
the radio sensitivity, although the specific form of galaxy 
evolution and dust properties also affected the radio-detectable 
sources. However, the large percentage 
($\sim 70$\%; see \S~\ref{secoverlap})
of \smm$>$5\,mJy blank-field counts recovered through radio 
selection shows that most of the bright submillimeter sources 
can be found this way, in agreement with predictions from the C02 
best-fit models.

Although microjansky radio sources have proved to be a crucial
resource for identifying and studying the nature of 
\sub-luminous galaxies, the \sub-detected samples are only a 
small fraction of the radio population. 
In this paper we present multiwavelength observations of a large
sample of microjansky radio sources detected in ultradeep 1.4\,GHz 
maps of the HDF-N, SSA13, and SSA22 fields to learn about the nature
and evolution with redshift of the microjansky radio population, 
as well as the implications of using this population to study 
the bright \sub\ source population. We assume a flat 
$\Omega_\Lambda=2/3$ Universe with $H_0=65$~km~s$^{-1}$~Mpc$^{-1}$.

\section{Observations}

\subsection{Radio imaging}
\label{secradio}

Data at 1.4\,GHz were obtained for the HDF-N, SSA13, and SSA22 
fields using the VLA. The HDF-N radio map 
(\markcite{richards00}Richards 2000) was observed in the A-array 
configuration, resulting in an $\sim 1.5''$ beam and reaching a 
sensitivity of 40~$\mu$Jy ($5\sigma$). The SSA13 radio map
(E.~B.~Fomalont et al., in preparation) was observed in both 
the A and B-array configurations for a total of 100~hours,
resulting in an $\sim 1.5$\arcsec\ beam and
reaching a sensitivity of 25~$\mu$Jy (5$\sigma$).
The SSA22 radio map (M.~Yun et al., in preparation) was observed 
in the B-array configuration for 12~hours, resulting in an 
$\sim 5''$ beam and reaching a sensitivity of $60~\mu$Jy 
($5\sigma$).

\subsection{Optical and near-infrared imaging}
 
Deep $I$-band imaging was used to identify the radio sources. 
The $I$-band image of the HDF-N field 
was obtained with the Subaru Prime Focus Camera (Suprime-Cam; 
\markcite{miya98}Miyazaki et al.\ 1998) on the 
8.2~m Subaru\footnote{The Subaru telescope is operated by the
National Astronomical Observatory of Japan} telescope.
Details of the observations and reductions can be 
found in \markcite{barger02}Barger et al.\ (2002), and
catalogs of all the galaxies and stars in the field
are presented in P.~Capak et al., in preparation.
The HDF-N optical image covers $24'\times 24'$ to $I=25.8$ ($5\sigma$).
In the HDF-N field, 357 radio sources have $I$-band measurements.

The $I$-band image of the SSA13 field 
(\markcite{barger01a}Barger et al.\ 2001a) was formed from a mosaic 
of overlapping images obtained with the Low Resolution Imaging 
Spectrograph (LRIS; \markcite{oke95}Oke et al.\ 1995) on the
10~m Keck\footnote{The W.~M.~Keck Observatory is operated as a
scientific partnership among
the California Institute of Technology, the University of California,
and NASA, and was made possible by the generous financial support of the
W.~M.~Keck Foundation} telecopes. 
In the SSA13 field, 145 radio sources have $I$-band measurements.

The $I$-band image of SSA22 was obtained with the CFH12K camera 
(\markcite{cuillandre00}Cuillandre et al.\ 2000) on the
3.6~m Canada-France-Hawaii Telescope (CFHT). The 8.1~hour 
integration reaches $I=26.1$ ($5\sigma$). Details 
are presented in S.~C.~Chapman et al., in preparation.
In the SSA22 field, 122 radio sources have $I$-band measurements.

For $I>21$ we used $3''$ diameter
aperture measurements corrected to total magnitudes
using an average median offset. For $I<21$ we used
isophotal magnitudes measured to 1\% of the peak
surface brightness of each galaxy. Isophotal magnitudes provide
a better measure of the total magnitudes in these generally
low redshift galaxies with extended angular sizes.

%
%
\begin{figure*}[htb]
\centerline{
\psfig{figure=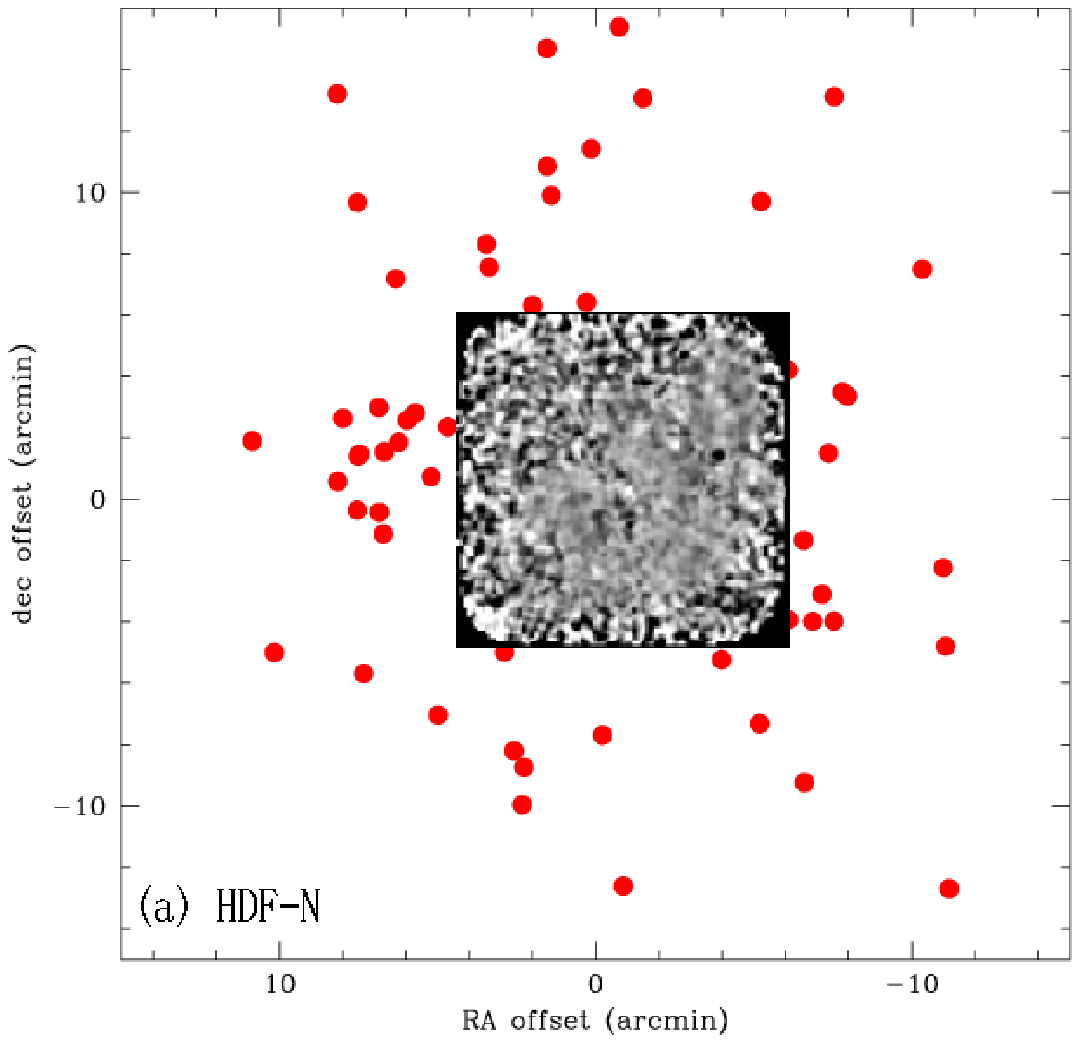,angle=0,width=2.5in}
\psfig{figure=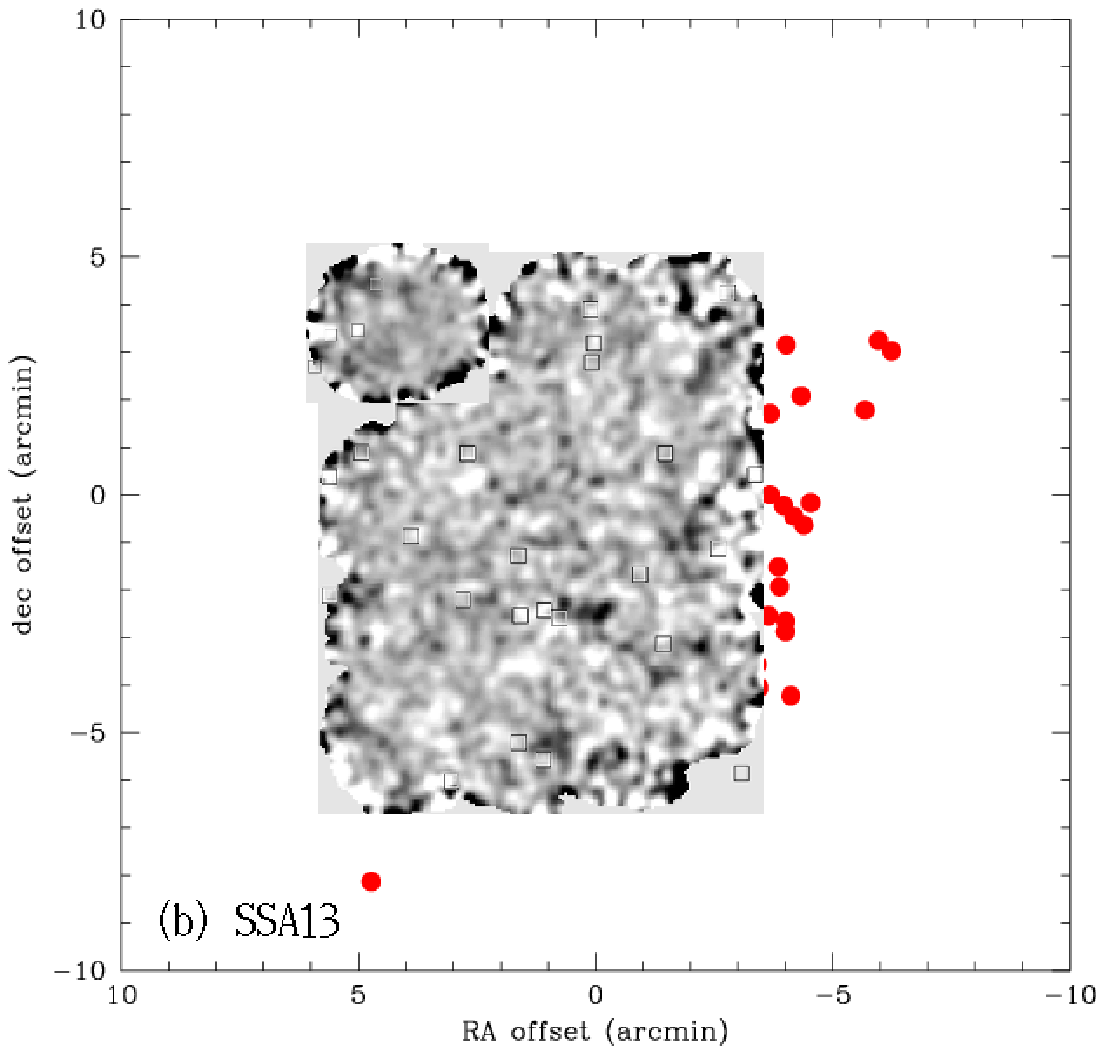,angle=0,width=2.5in}
\psfig{figure=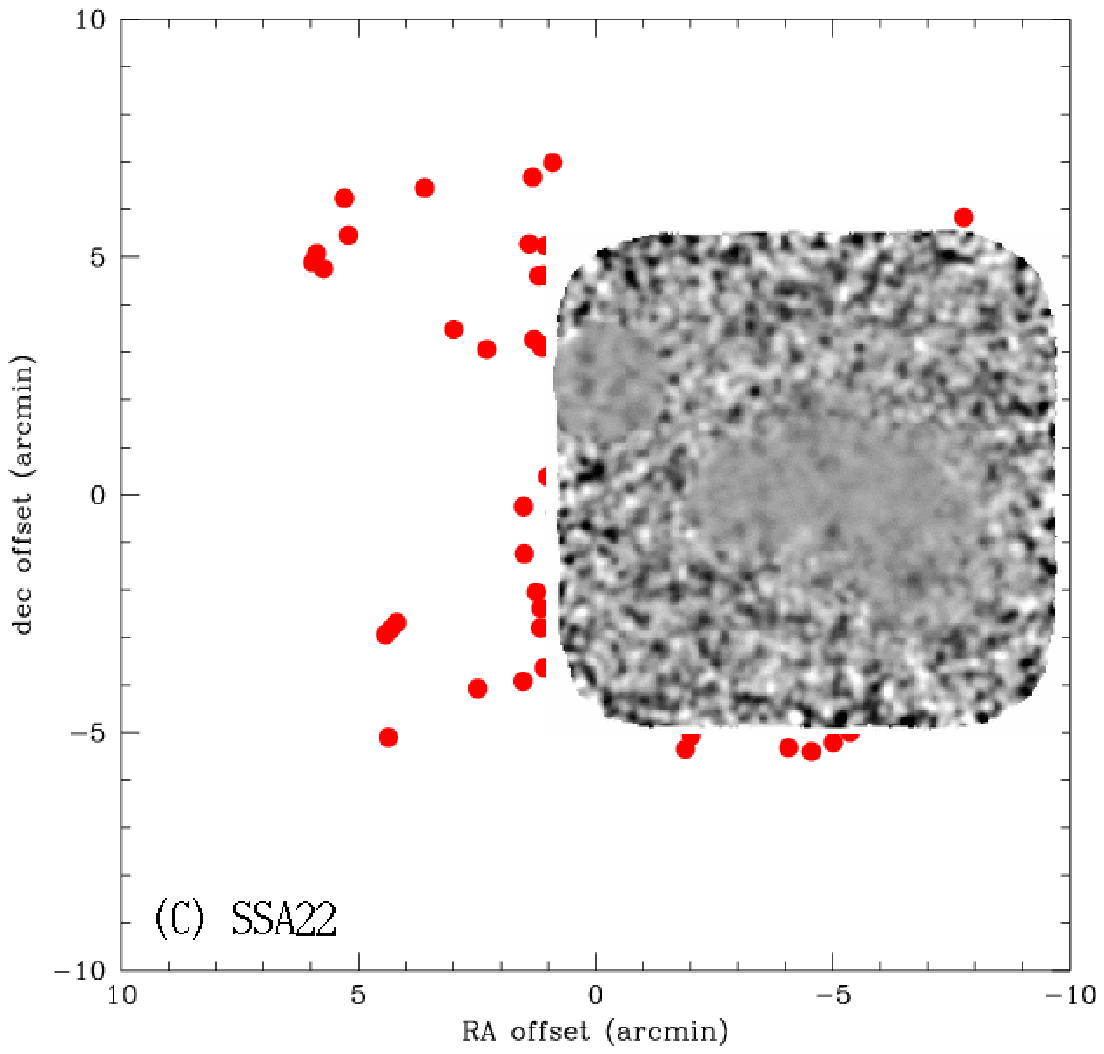,angle=0,width=2.5in}
}
\figurenum{1}
\caption{
Configuration of the submillimeter data in the HDF-N, SSA13, and SSA22
fields (as indicated in panels a,b,c).  
A central area of $\sim 120$ arcmin$^2$ has been mapped with
SCUBA to an uniform depth of $\sim 2$\,mJy in all three fields,
although small subregions go much deeper.
For illustration, we have indicated the radio sources with squares
within the field with the deepest radio map (SSA13).
Targeted submillimeter follow-up of $I>23.5$ radio sources was undertaken
outside of the SCUBA mapped regions, but within the VLA primary
beam (shown with filled circles), as well as within the SCUBA mapped regions
to confirm map detections.
The half-power point of the VLA primary beam at
1.4\,GHz is 30\arcmin\ diameter.
}
\label{fig1}
\addtolength{\baselineskip}{10pt}
\end{figure*} 

Deep $K$-band imaging of SSA22 was obtained
at UKIRT using the UKIRT Fast Track Imager (UFTI).
The small UFTI field ($50''\times 50''$) was centered on
the positions of the $>3\sigma$ submillimeter-detected radio sources.
Each source was imaged for a total of 3600~s using individual exposures 
of 60~s; the limiting magnitude in a $2''$ diameter 
aperture is $K=20.8$ ($5\sigma$). The fast tip/tilt, adaptively-corrected 
imaging resulted in seeing better than the median conditions of
$0.45''$ FWHM. The data were reduced using the Starlink UKIRT/UFTI
image processing tools under the {\sc oracdr} environment 
(\markcite{bridger00}Bridger et al.\ 2000). 
We wrote custom {\sc oracdr} scripts to optimize point-source
sensitivity in our essentially blank-field observations, creating 
flat-fields from each 9-point dither and high signal-to-noise thermal
background images from 60~minutes of data.

For sources in the HDF-N, $K$-band magnitudes were inferred from the 
deep $HK'$ image of the Hubble Flanking Fields 
(\markcite{barger99}Barger et al.\ 1999; $HK'-K=0.3$)
obtained using the University of Hawaii Quick Infrared Camera (QUIRC; 
\markcite{hodapp96}Hodapp et al.\ 1996) on the 2.2~m University of 
Hawaii telescope and the CFHT. The image reaches an effective 
limiting magnitude of $K=20.1$ ($5\sigma$).

\subsection{Optical spectroscopy}

Spectroscopic measurements of 158 radio sources in the HDF-N and 51 
in the SSA13 field were made with LRIS on Keck
and with the HYDRA spectrograph 
(\markcite{barden94}Barden et al.\ 1994) on the 3.5~m WIYN telescope.
{\it We hereafter refer to the radio sources in the HDF-N and SSA13 
fields with spectroscopic measurements as our spectroscopically-observed 
radio sample.} Redshift identifications were obtained for 169 sources:
126 in the HDF-N and 43 in SSA13.
The spectral observations and reductions are presented and described 
in \markcite{bcr00}BCR00 and E.~B.~Fomalont et al., in preparation
(see also \markcite{cohen00}Cohen et al.\ 2000 and
\markcite{barger02}Barger et al.\ 2002 for many of the redshifts 
in the HDF-N field).

\subsection{Submillimeter imaging}
\label{secsmm}

The submillimeter data were taken in SCUBA jiggle map, SCUBA 
raster map, or SCUBA photometry mode at 850\mum during a number of 
observing runs with good to excellent conditions. 
The configuration of the data in all three fields is similar:
a roughly $10'\times 10'$ region in the center of the 
40\arcmin\ diameter VLA primary beam was mapped with SCUBA in
either jiggle map or raster map mode to an average sensitivity of
2\,mJy r.m.s.~(with small subregions reaching as deep as 0.5\,mJy r.m.s.).
These maps present an unbiased study of the submillimeter
properties of the microjansky radio sources.
The SCUBA jiggle maps are described in detail in
\markcite{bcs99}Barger, Cowie, \& Sanders (1999), BCR00, and
\markcite{barger01a}Barger et al.~(2001a, b), and
the SCUBA raster maps are described in Borys et al.~(2002) and
C.\ Borys et al., in preparation. For the jiggle and raster maps,
submillimeter fluxes were measured at the positions of the 1.4\,GHz
($>5\sigma$) sources from the catalogs of
\markcite{richards00}Richards (2000; HDF-N),
E. B. Fomalont et al., in preparation (SSA13), and
S.~C.~Chapman et al., in preparation (SSA22) using beam-weighted 
extraction routines that include both the positive and negative 
portions of the beam profile. 

Additional targeted SCUBA photometry mode observations were made
of the majority of the optically-faint radio sources over a wider part of the 
VLA primary beam, including some sources in regions already mapped by 
SCUBA. The photometry data are 
described in C01, C02, and S.~C.~Chapman et al., in preparation. 
These observations were centered on the radio source coordinates
producing a flux measurement at that position. The configuration of 
the \sub\ data in each of the three fields is depicted in Fig.~\ref{fig1}.
Of the 278  radio sources in the three fields with submillimeter
measurements, all have 850\mum flux measurements with 
uncertainties less than 5\,mJy, and of these, 219 have uncertainties 
less than 2.5\,mJy. In total, we significantly ($>3\sigma$) detect 
in the submillimeter 50 of the 278 radio sources. 
Twenty-eight of these lie in the jiggle or raster maps, and 22
are from photometry mode. For reference, there is about one blank
sky SCUBA source with ${>}\,5\,$mJy for every 200 SCUBA beams.

\section{Optical properties and spectroscopic redshift distribution
of the microjansky radio sources}
\label{secz}

\markcite{richards99}Richards et al.\ (1999) presented the overall 
$I$ magnitude distribution for the 1.4\,GHz and 8.4~GHz samples 
in the HDF-N and the 8.4~GHz sample in the SSA13 field. They
found that the optical counterparts had a wide range of
magnitudes with a median $I=22.1$. The present deeper
$I$-band observations remain consistent with this result. The 
median $I$ for the HDF-N and SSA13 1.4\,GHz sources with
$I$-band measurements is $22.37^{+0.11}_{-0.18}$ (68\% confidence).
The distribution implies that the bulk of the radio sources 
(just over 60\%) should be bright enough ($I\lesssim 23.5$)
for spectroscopic identification. The remaining sources will mostly
be too optically faint for spectroscopic identification and 
will require another method for determining their redshifts.

%
%
\begin{inlinefigure}
\centerline{
\psfig{figure=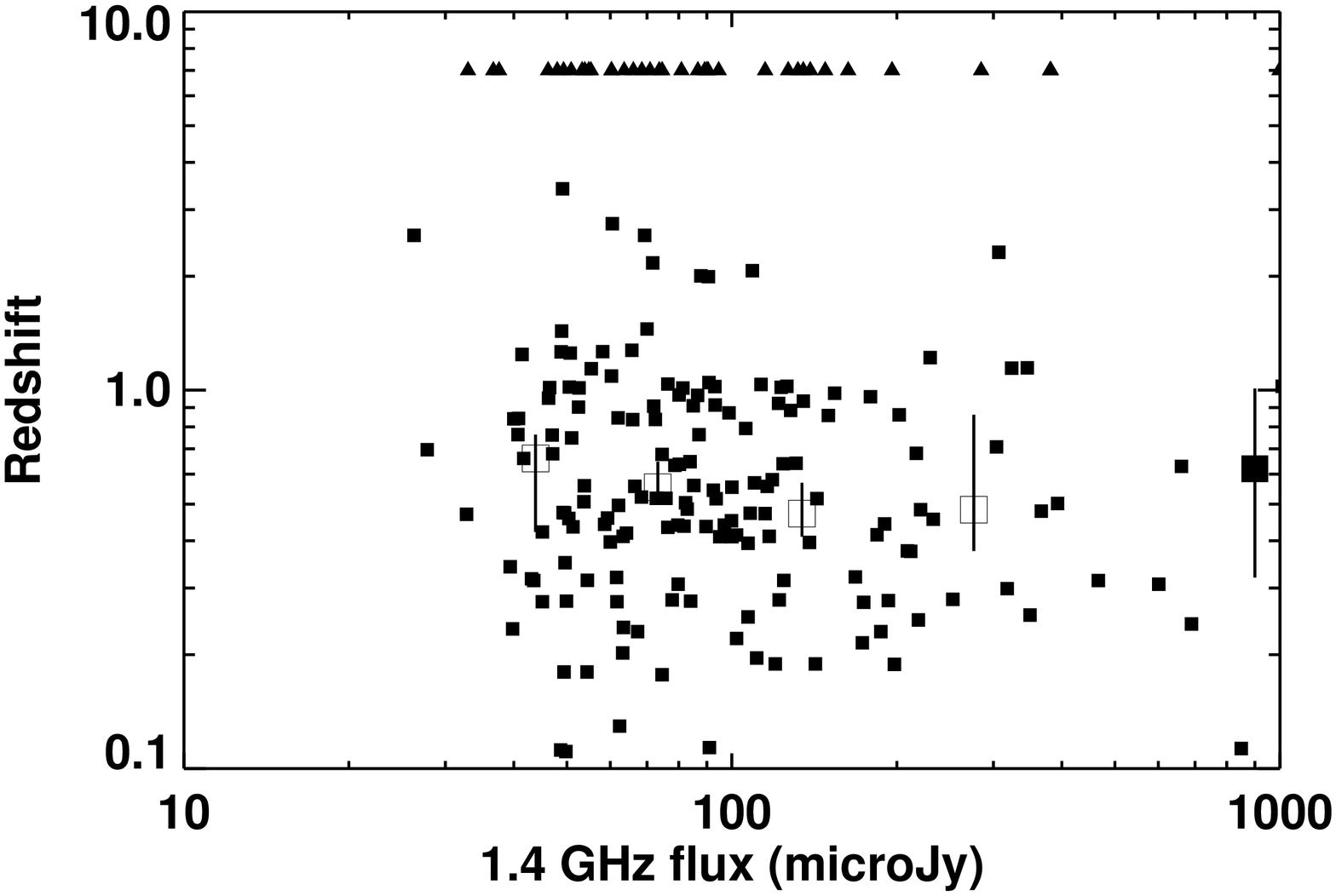,angle=90,width=3.5in}
}
\figurenum{2}
\caption{Redshift versus 1.4\,GHz flux for the spectroscopically-observed
radio sources in the HDF-N and SSA13 fields. The solid squares
denote sources with redshift identifications, and the solid
triangles at $z=7$ denote sources for which no redshifts could
be obtained. The large open squares denote the median redshifts
for the identified sources. The 68\% confidence range on
the median redshifts are indicated by the error bars.
The large solid square with error bars shows the median
redshift and 68\% confidence range of the millijansky
radio sample of Waddington et al.~(2001).
\label{fig2}
}
\addtolength{\baselineskip}{10pt}
\end{inlinefigure}

In Fig.~\ref{fig2} we plot redshift versus \rad\ for the
spectroscopically-observed radio sample. Sources with redshifts are
denoted by solid squares, and sources for which no redshifts could
be obtained are denoted by solid triangles at $z=7$.
The large open squares denote the median redshifts for the identified
sources. The median redshift distribution
is essentially flat across the entire microjansky regime probed.
We also plot the median redshift (large solid square)
from a millijansky radio study by
\markcite{waddington01}Waddington et al.~(2001).
This point shows a continuation of the flat redshift distribution
to brighter radio fluxes.
The selection of optically-bright sources inherent in a
spectroscopically-identified sample means we are
likely missing the high redshift tail of the distribution
function. This tail could be skewing to higher redshifts as we
move to fainter radio fluxes while the main body of the redshift
distribution remains invariant. We shall return to this
point when we discuss the submillimeter properties of the
radio sources.

In Fig.~\ref{fig3} we plot rest-frame absolute $I$-band 
magnitude, $M_I$, calculated using a standard spiral galaxy
$K$-correction, versus redshift for the 
spectroscopically-identified radio sources (solid squares).
The effect of the $I\sim 23.5$ spectroscopic limit on $M_I$ is indicated 
by the dashed line, and the effect of our growing incompleteness beyond 
$I\sim 22$ is indicated by the dotted line. 
Fig.~\ref{fig3} shows a remarkably narrow 
range in $M_I$ with redshift (as indicated by the solid lines) that
suggests the radio sources are chosen from approximately $L^*$ 
optical galaxies, regardless of redshift, and the radio hosts are, 
at a crude level, standard candles in their optical properties. 
In the redshift range $z=0.1$--0.5 the mean $M_I=-22.0\pm0.9$; for
$z=0.5-0.8$ the mean $M_I=-22.6\pm0.8$ (by contrast, classical radio
galaxies are approximately 5--10$\times L^*$ optical galaxies). As can be 
seen from Fig.~\ref{fig3}, the small increase in absolute magnitude 
at higher redshifts largely may be due to the apparent magnitude 
selection effects.

To verify that this result is not an effect of our flux limited radio 
sample, we construct a toy model whereby a Schechter luminosity function
with $M_I^*=-22.5$ is evolved in luminosity, increasing with redshift
as (1+$z$)$^3$ from $z=0.1-1.5$, and normalized by the volume
element, $d$V.
If we were to assume that radio luminosity is proportional to optical ($I$-band)
luminosity, our observed range in $M_I$ would result in only
sub-$L^*$ radio sources lying in our sample.
However, if we apply our radio flux limit ($\sim$40$\mu$Jy) to the model, 
we notice that 
we can detect only those radio sources with luminosities greater than $L^*$
at all redshifts $>0.3$, with 70\% of the detectable sources in the
observed volume lying outside of the equivalent $M_I$ range.
We conclude that our observed result (microjansky radio sources being hosted
by similarly luminous optical galaxies) is not a selection effect of our
flux limited radio sample.

%
%
\begin{inlinefigure}
\centerline{
\psfig{figure=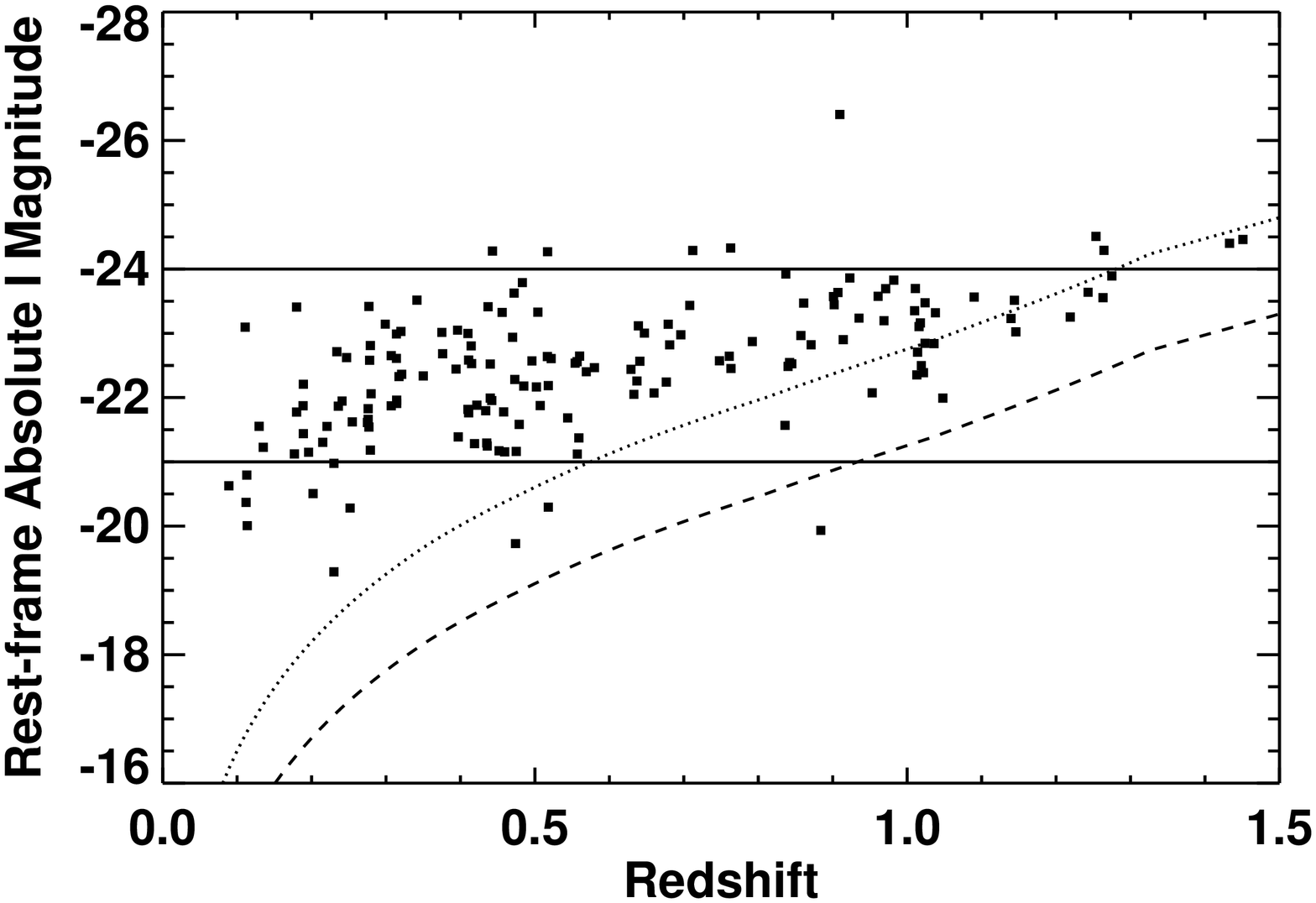,angle=90,width=3.5in}
}
\figurenum{3}
\caption{
$M_I$ versus redshift for the radio sources in the HDF-N and SSA13
fields with spectroscopic
redshifts (solid squares). The solid lines show the approximate
range in $M_I$ from $-21$ to $-24$ where the majority
of the sources lie. The dashed line illustrates the effect of an
$I\sim 23.5$ spectroscopic limit on $M_I$, and the dotted line
illustrates that of an $I\sim 22$ selection.
\label{fig3}
}
\addtolength{\baselineskip}{10pt}
\end{inlinefigure}

It is not clear why the radio sources should all lie in high
optical luminosity galaxies. At the higher redshifts the restricted 
absolute magnitude range can be understood from the selection effects: 
the lower cut-off is due to low optical luminosity galaxies being faint 
and not spectroscopically identified, and the upper cut-off reflects the 
small number of galaxies with optical magnitudes above the optical $L^*$. 
However, at the lower redshifts there is no selection against low
optical luminosity galaxies, and the absence of these galaxies is striking. 
Many of the low optical luminosity galaxies at low redshifts have blue 
colors and substantial ongoing star formation that might be expected
to produce substantial synchrotron emission; thus, we might
expect this population to correspond to a large part of the
radio population. The absence of these objects argues that
they are weak in the radio relative to their optical
(and ultraviolet) luminosities.

This presumably is the same effect that is seen in the local 
far-infrared (FIR)-radio correlation, which is not precisely 
linear; low optical luminosity galaxies have lower radio 
luminosities than expected (\markcite{fitt88}Fitt et al.~1988;
\markcite{cox88}Cox et al.~1988;
\markcite{dev89}Devereux \& Eales 1989).
One likely explanation is that small galaxies are not very 
effective radio sources because cosmic rays are more likely to escape 
by diffusion or convection (\markcite{chi90}Chi \& Wolfendale 1990;
\markcite{condon91}Condon et al.\ 1991.)
As our sources are selected from a flux-limited radio survey,
low optical luminosity galaxies will be eliminated
preferentially from our sample. This is a serious problem for 
radio estimates of the local star formation rate density
(e.g., \markcite{mobasher99}Mobasher et al.\ 1999;
\markcite{haarsma00}Haarsma et al.\ 2000), as
a substantial fraction of the ultraviolet luminosity density is
generated by faint sub-$L^*$ galaxies at low redshifts
(e.g., \markcite{csb99}Cowie, Songaila, \& Barger 1999;
\markcite{wilson02}Wilson et al. 2002). The radio-based
star formation rate density estimates already tend to lie at the 
high end of the values measured in the $z=0$--1 range, but despite
this, the lowest redshift values must be substantially underestimated. 

\section{Submillimeter properties of the microjansky radio sources}
\label{secprop}

In Fig.~\ref{fig4} we plot \smm\ versus $I$-band magnitude for the
radio sources with submillimeter measurements,
and in Fig.~\ref{fig5} we plot \smm\ versus \rad.
The data are divided both according to observation technique
(triangles for jiggle map or raster mode, circles for photometry mode) 
and according to submillimeter detection significance (large solid 
symbols for $>3\sigma$, small open symbols for $<3\sigma$).
Analyses of submillimeter source reliability from
SCUBA maps suggest a threshold of $4\sigma$
(\markcite{eales00}Eales et al.\ 2000;
\markcite{scott02}Scott et al.\ 2002;
\markcite{borys02}Borys et al.\ 2002;
\markcite{cowie02}Cowie, Barger, \& Kneib 2002);
thus, we have enclose our $>4\sigma$ sources in a second larger symbol. 
However, confidence
in a targeted \sub\ detection of a known optically-faint radio source
is much higher than in a randomly detected \sub\ source. For the
278 radio sources with submillimeter measurements, we expect only
0.5 spurious $3\sigma$ sources.

%
%
\begin{inlinefigure}
\psfig{figure=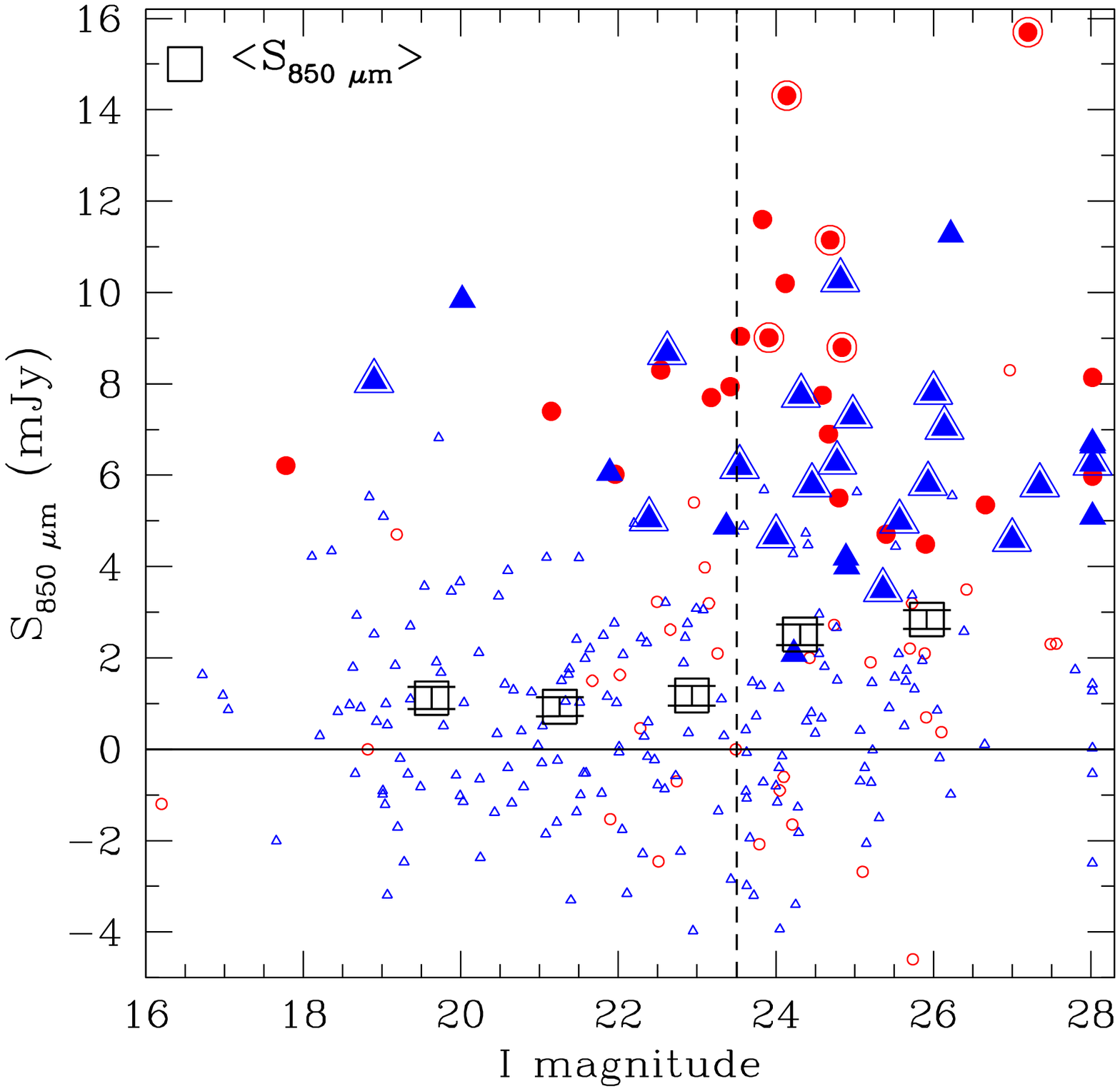,angle=0,width=3in}
\vspace{6pt}
\figurenum{4}
\figcaption{
Submillimeter flux versus $I$-band magnitude
for the microjansky radio sources in the HDF-N, SSA13, and SSA22 fields
with submillimeter measurements. Circles denote SCUBA photometry
observations and triangles denote SCUBA jiggle map or raster map
observations. Large filled symbols denote sources with $>3\sigma$
submillimeter fluxes. Sources with $>4\sigma$ submillimeter fluxes
are enclosed in a second larger symbol. The error-weighted mean
submillimeter fluxes are shown as large open squares
with $1\sigma$ uncertainties.
The vertical dashed line at $I=23.5$ shows our division between
optically-faint and optically-bright sources. Error bars have
not been plotted but are typically ${\sim}\,2\,$mJy.
\label{fig4}
}
\addtolength{\baselineskip}{10pt}
\end{inlinefigure}

\subsection{Statistical analyses}

In Fig.~\ref{fig4} we show the error-weighted mean submillimeter
fluxes (open squares with $1\sigma$ uncertainties) for five bins
of 53 sources at successively fainter $I$-band magnitudes.
The sample as a whole has an error-weighted mean \smm$=1.73\pm 0.09$\,mJy.
Since the distribution of the noise is not Gaussian
(see \markcite{cowie02}Cowie, Barger, \& Kneib 2002), the computed 
error may underestimate the true value. However, when Monte Carlo
simulations of an equal number of randomly distributed sources are 
analyzed in the same way, the high significance of the result is
confirmed. The optically-faint ($I>23.5$) subsample
is highly significant at $2.37\pm 0.13$\,mJy, while the 
optically-bright ($I<23.5$) subsample is also significant at
$1.09\pm 0.13$\,mJy. Thus, the optically-faint radio
sources are about two times more submillimeter luminous, on average,
than the optically-bright radio sources. While we have
somewhat oversampled the optically-faint radio sources with our
targeted submillimeter photometry observations,
the result is consistent when we consider only our contiguous
submillimeter jiggle map regions.

Both the optically-faint and optically-bright radio-selected samples
have higher submillimeter fluxes than X-ray-selected samples:
for the 1~Ms {\it Chandra} Deep Field-North X-ray sample
the error-weighted mean \smm$=1.21\pm 0.14$\,mJy 
(\markcite{barger01b}Barger et al.\ 2001b), and for the 75~ks
Elais-N2 sample, $1.25\pm 0.40$\,mJy
(\markcite{almaini02}Almaini et al.\ 2002). In fact, 
\markcite{barger01}Barger et al.\ (2001b) found that the submillimeter
signal of their X-ray sample was largely driven by the subset of
sources that overlapped with optically-faint radio sources.

%
%
\begin{inlinefigure}
\psfig{figure=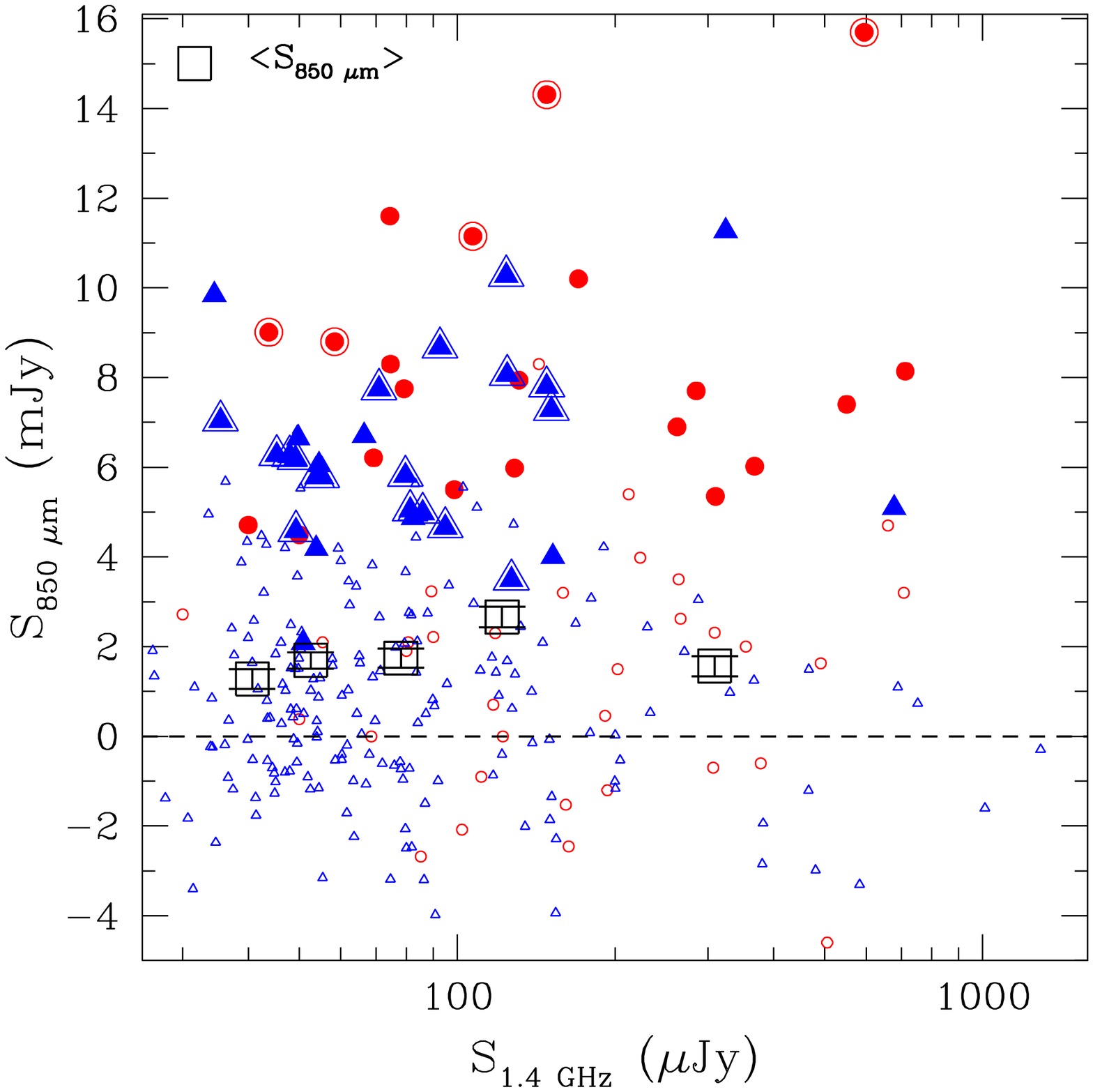,angle=0,width=3in}
\vspace{6pt}
\figurenum{5}
\caption{
Submillimeter flux versus radio flux for the microjansky radio sources
in the HDF-N, SSA13, and SSA22 fields with submillimeter measurements.
Symbols are as in Fig.~\ref{fig4}.
\label{fig5}
}
\addtolength{\baselineskip}{10pt}
\end{inlinefigure}

In Fig.~\ref{fig5} we show the error-weighted mean submillimeter
fluxes (open squares with $1\sigma$ uncertainties)
at successively fainter 1.4\,GHz fluxes. The mean rises (from
\smm$\sim 1$\,mJy to 3\,mJy) with increasing radio flux out to about
100\,$\mu$Jy before dropping to \smm=1.2\,mJy for the brightest
radio sources.  For radio fluxes approaching the millijansky regime,
we expect an increasing contribution from active galactic nuclei (AGN).
This appears to be reflected in the trend of the means:
the brightest radio sources in our sample are, on average, not the 
strongest \sub\ emitters (luminous
AGN typically have low \ratio\ ratios). Since the SSA13 radio image
is deeper than the HDF-N and SSA22 images, the number of radio
sources probed in the submillimeter is not simply proportional to the
radio source counts. However, the statistics should not be affected
by this, unless significant large-scale structure dominates the
small field covered to this depth. Note that within the area probed
by our \sub\ observations, only two radio sources have radio fluxes
higher than 1\,mJy.

\subsection{Direct submillimeter source detections}
\label{secbright}

Fifty of the radio sources in our sample are significantly
($>3\sigma$) detected individually in the submillimeter.
Of these, 38 fall into the optically-faint ($I>23.5$) region of 
Fig.~\ref{fig4}, which is consistent with the conclusion of previous 
SCUBA surveys that most submillimeter sources are optically faint
(e.g., BCR00; \markcite{smail02}Smail et al.\ 2002; C01).
The percentage of \smm$>5$\,mJy ($>3\sigma$) submillimeter sources 
detected in our survey versus optical brightness is presented in Table~1;
the radio selection technique becomes
highly efficient for optical magnitudes fainter than $I\sim24$.

While the submillimeter detection rate of $I<23.5$ radio sources
is rather poor, this population still makes up a fraction
of our submillimeter sample. To quantify this, we consider
only regions of our survey that were mapped with SCUBA to avoid the
bias introduced by the optically-faint criterion used for the targeted
photometry follow-up. Using our combined jiggle and raster maps 
of the three fields, we find that 20 of the submillimeter-detected 
sources have $I>23.5$ and 9 have $I<23.5$. Thus, $69\pm9$\%,
where the errors correspond to the 1 sigma variance in the expectation value,
of the radio-\sub\ population has $I>23.5$.
This appears to be inconsistent
with the claim by \markcite{ivison02}Ivison et al.\ (2002), based 
on a smaller sample, that the bulk of the radio-detected submillimeter
sources are optically bright. 

\subsection{Overlap of the bright submillimeter population with the
microjansky radio population}
\label{secoverlap}

Using our jiggle and raster mapped regions ($>300$~arcmin$^2$), we can
determine what fraction of bright \sub\ sources are picked out by
an ultradeep radio sample. As the radio properties in the SSA22 field
are highly non-uniform due to bright radio sources, we again restrict 
our analysis to the HDF-N and SSA13 fields. We also restrict 
to $>4\sigma$ sources with \smm$>5$\,mJy.
We have adopted an 8\arcsec\
search radius to avoid missing associations due to the positional
uncertainties of the S/N$\sim$4 SCUBA detections.
Note that the 8\arcsec\ search radius should not give rise to any
spurious associations as the surface density of radio sources is
only $\sim2$~arcmin$^{-2}$ in the central regions of our maps.
We find that $66\pm7$\% of bright submillimeter sources
(29 of the /N$\sim$4 SCUBA detections)
have radio identifications. This is consistent with
\markcite{ivison02}Ivison et al.\ (2002),
who found that 18 of the 30 submillimeter sources in their sample
have radio identifications, which translates to a percentage of
$60\pm9$\%. From \S~\ref{secbright}, about 70\% of these
counts will be picked out by the optically-faint radio sources.

%
%
\begin{inlinefigure}
\psfig{figure=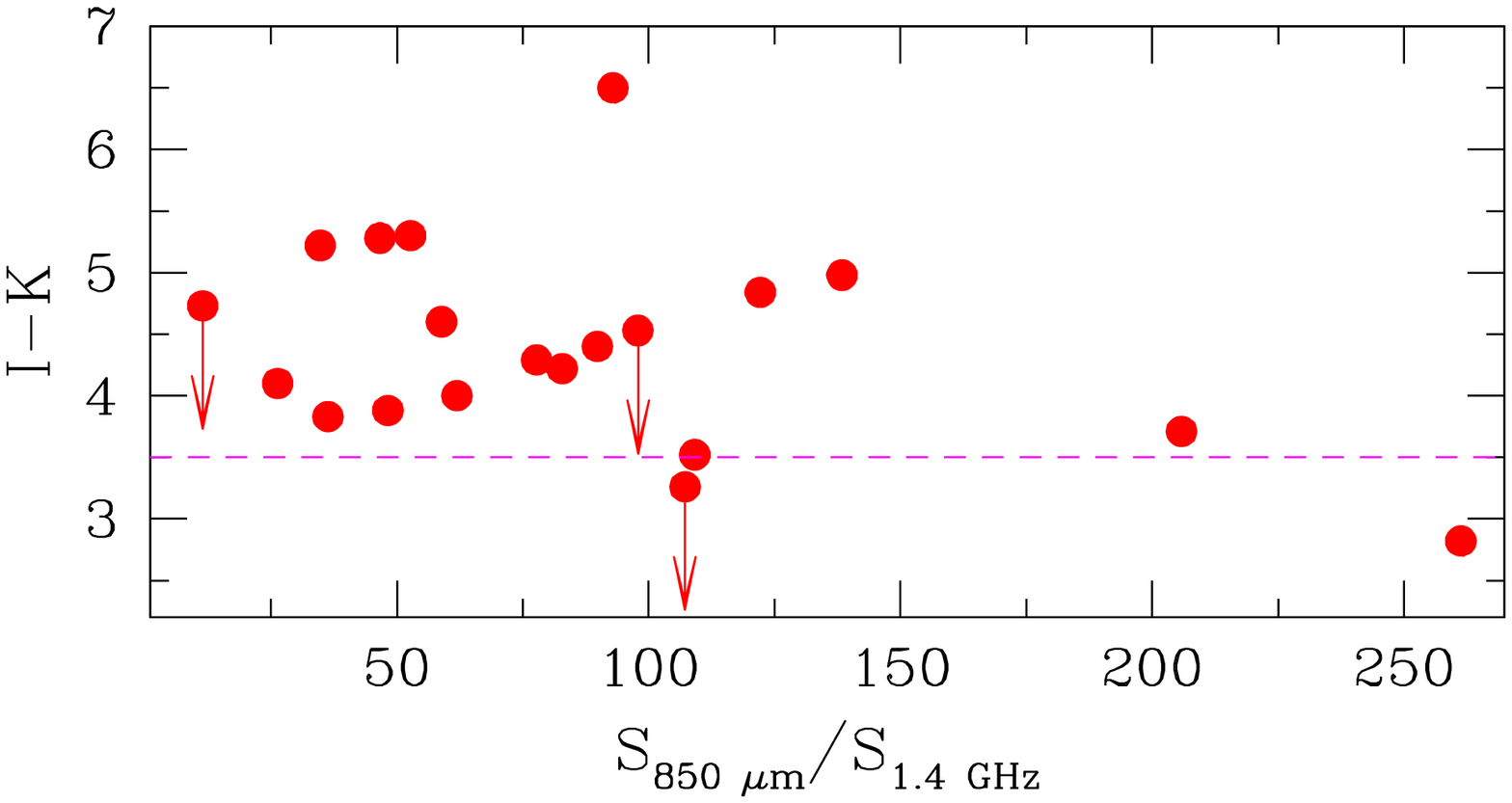,angle=0,width=3.5in}
\vspace{6pt}
\figurenum{6}
\caption{
$I-K$ versus the radio of submillimeter to radio
flux for the microjansky radio sources in the
HDF-N and SSA22 fields with both submillimeter and $K$-band measurements.
Sources with marginal $K$-band detections are shown as $3\sigma$ upper
limits. The dashed line shows the Very Red Object limit, $I-K=3.5$.
\label{fig6}
}
\addtolength{\baselineskip}{10pt}
\end{inlinefigure}

\subsection{$I-K$ Colors}
\label{secik}

We have $K$-band measurements for a subset of our radio sample.
Both the optically-bright and optically-faint radio subsamples
have significantly redder $I-K$ colors than the field galaxy population.
The optically-faint radio sources show a monotonic trend to redder 
colors (\markcite{richards99}Richards et al.\ 1999; BCR00):
at $I=22$, median $I-K=2.9$ and at $I=24.5$, median $I-K=3.4$.
For comparison, the field galaxy sample has median $I-K\sim2$ at
$I\sim22$ and median $I-K\sim2.3$ at $I\sim24.5$
(e.g., \markcite{barger99}Barger et al.\ 1999).

In Fig.~\ref{fig6} we plot $I-K$ versus \ratio\ for the $3\sigma$
\sub-detected radio sources with $K$-band measurements. All of the
sources have $I$-band detections, but three of the sources are only
marginally detected at $K$ and are shown as $3\sigma$ upper limits.
The \sub-detected sources have median $I-K=4.3$ and median
$K=20.7$. From Fig.~\ref{fig6} we can see that a color cut of
$I-K>3.5$ includes 95\% of the \sub-detected sources.
This is consistent with the results of
\markcite{wehner02}Wehner, Barger, \& Kneib (2002), who used
a submillimeter survey of three massive cluster fields to determine
that the median submillimeter source has $K=20-20.3$,
depending on the cluster sample used, and that
$I-K>3.5$ sources (Very Red Objects) contribute about two-thirds of
the 850\mum extragalactic background light.

There is no obvious trend of $I-K$ color with \ratio\
in Fig.~\ref{fig6}, implying that the near-infrared colors
of the submillimeter-selected galaxies are insensitive
to redshift over the $z=1$--3 range as would be expected if the
red colors are primarily caused by dust extinction. 
\section{Millimetric redshift estimation for the microjansky 
radio-selected sample}
\label{seccy}

\subsection{Evolution of the radio population with redshift}
\label{radioev}

The radio power of the spectroscopically-identified sample
in the HDF-N and SSA13 fields is shown in Fig.~\ref{fig7}, 
where we have computed the radio 
$K$-correction assuming a synchrotron spectral index of $-0.8$. 
The selection limits of the HDF-N and SSA13 samples 
(40~$\mu$Jy and 25~$\mu$Jy, respectively) are shown by the solid curves. 
The conversion to FIR luminosity, computed using the local FIR-radio 
correlation (e.g., \markcite{condon92}Condon 1992), is shown
on the right-hand axis, and the dotted and dashed lines
show the FIR luminosity limits for luminous infrared galaxies (LIGs) 
and ultraluminous infrared galaxies (ULIGs), respectively.

There is a rapid evolution in the upper bound of the radio power. 
This is the well-known effect that large galaxies at $z\sim1$ 
frequently have very large star formation rates while at lower 
redshifts the rates turn down rapidly
(\markcite{cowie96}Cowie et al.\ 1996, where the phenomenon is 
called downsizing). An equivalent description is that near-ULIGs or 
ULIGs are extremely rare at low redshifts, but the
number density rises rapidly with redshift so that they are relatively 
common by $z=1$ (e.g., \markcite{blain99}Blain et al.\ 1999; 
BCR00). In combination with the lower-bound selection imposed 
by the flux limits, this effect means that we have a rapidly changing 
but rather well-defined population with redshift. 
At low redshifts ($z<0.4$) we are selecting Milky Way type galaxies,
at intermediate redshifts we are selecting LIGs, and at $z>1$
we are selecting only near-ULIGs or ULIGs.

%
%
\begin{inlinefigure}
\centerline{
\psfig{figure=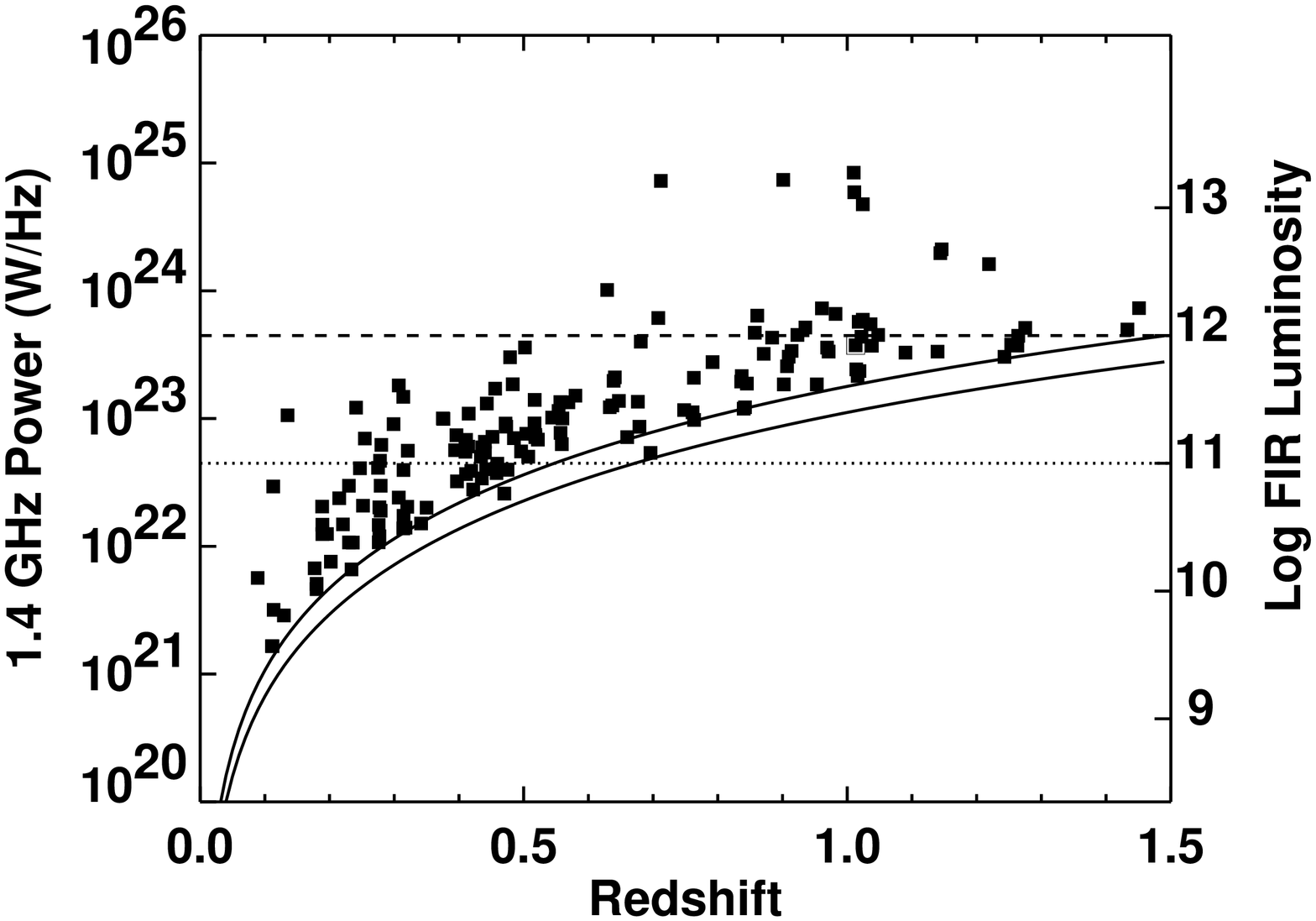,angle=90,width=3.5in}
}
\figurenum{7}
\caption{
Radio power versus redshift for the radio sources
in the HDF-N and SSA13 fields with spectroscopic redshifts.
The solid curves show the selection limits for the two
fields (40~$\mu$Jy and 25~$\mu$Jy, respectively). The right
vertical axis shows the logarithm of the FIR luminosity (in solar
luminosities) that would be inferred from the radio power, based
on the local FIR-radio correlation. The FIR luminosity
limits of LIGs ($10^{11}~L_\odot$) and ULIGs ($10^{12}~L_\odot$)
are shown by the dotted and dashed lines, respectively.
\label{fig7}
}
\addtolength{\baselineskip}{10pt}
\end{inlinefigure}

\subsection{Average \ratio\ flux ratios}
\label{avrat}

Our present dataset enables us to determine the normalization with 
redshift of the error-weighted mean \ratio\ flux ratio.
In Fig.~\ref{fig8} we plot the logarithm of the error-weighted mean
\ratio\ for the radio sources with spectroscopic
redshifts and submillimeter measurements (large open squares).
The four $>3\sigma$ submillimeter-detected sources with
redshifts (see \S~\ref{hizids}) are shown as dots.
We include on the figure the tracks with redshift for three
spectral energy distributions (SEDs) with different dust properties:
Arp~220 (short dashed), a typical {\it IUE} starburst (solid),
and a Milky Way type galaxy (dotted). We have calculated these tracks
using the \markcite{dale01}Dale et al.\ (2001) and 
\markcite{dale02}Dale \& Helou (2002) empirical
SED library. We also include three tracks from the literature
(DCE00; CY00; \markcite{yc02}Yun \& Carilli 2002) that were
derived using different galaxy samples and assumptions; these tracks
are the most widely used in millimetric redshift estimation.
As would be expected from the luminosity selection, the
error-weighted means follow the track of a quiescent
to moderately star-forming galaxy at lower redshifts,
while at higher redshifts the observations migrate to
values $1\sigma$ below the Arp~220 and CY00 tracks.
(We note that \markcite{yun01}Yun, Reddy, \& Condon (2001) found
that sources like Mrk~231, whose AGN has a radio luminosity 
comparable to its starburst luminosity, likely constitute a few to 
10\% of the \sub\ population. Such sources may lie
preferentially in our radio-selected sample and hence have lowered 
our high-redshift ULIG points by about 0.3 in dex.)

%
%
\begin{inlinefigure}
\psfig{figure=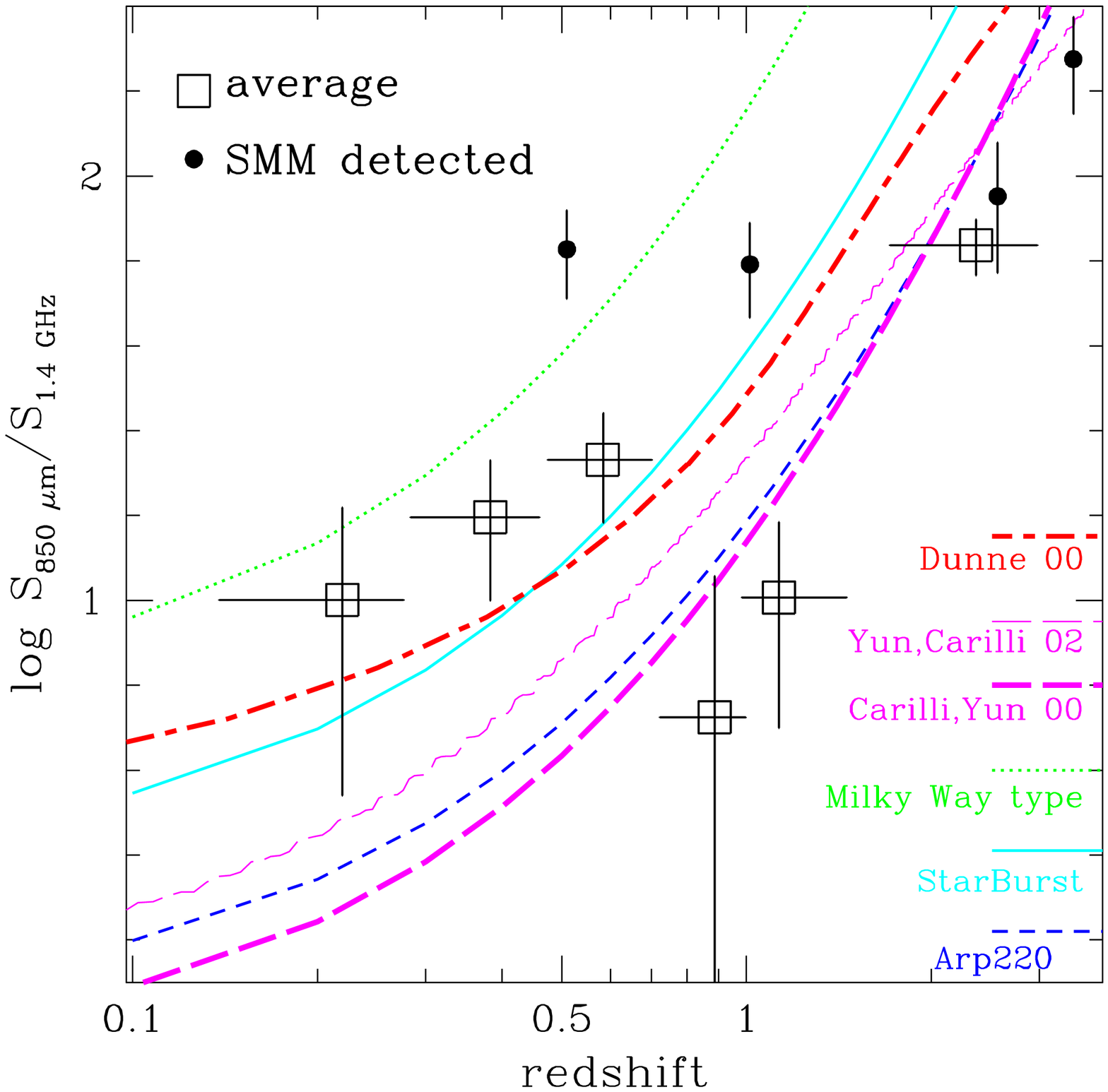,angle=0,width=3in}
\vspace{6pt}
\figurenum{8}
\caption{
$\log$ \ratio\ versus redshift for the microjansky radio sources with
spectroscopic redshifts and submillimeter measurements.
Squares denote the error-weighted means in six equal-number bins.
The four $>3\sigma$ \sub-detected sources with reliable redshifts
are shown as dots. The tracks are Arp~220, a Milky Way type galaxy,
a typical {\it IUE} starburst, and
Dunne et al.\ (2000), Carilli \& Yun (2000), and Yun \& Carilli (2002).
}
\label{fig8}
\addtolength{\baselineskip}{10pt}
\end{inlinefigure}

The radio power and the FIR luminosity correlate locally with dust 
temperature, and there is no evidence that this relation varies 
significantly with redshift 
(\markcite{chapman02c}Chapman et al.\ 2003, \markcite{bbc02}Blain, Barnard \&
Chapman 2003). Given this,
the evolution in radio power and in \ratio\ with redshift show that
for the microjansky radio population, millimetric redshift estimates
at low redshifts are best made with a template intermediate
between a Milky Way type template and a starburst template,
and those at high redshifts ($z>1$) with an Arp~220 template.
Thus, we can apply the millimetric redshift technique to other
radio-\sub\ populations {\it provided} we have knowledge of the
characteristic FIR luminosity or redshift of the population
before we choose a template SED.

In Fig.~\ref{fig3} we showed that nearly all radio sources with
redshifts have similar $M_I$. This suggests that the optically-faint 
radio sources, rather than being low optical luminosity galaxies at 
moderate redshifts are instead luminous galaxies at $z>1$, where the 
negative $K$-correction in the \sub\
provides a nearly one-to-one mapping of \sub\ flux to \sub\
luminosity. For the sources that are individually detected in the \sub,
the \smm\ fluxes (see Fig.~\ref{fig4}) give FIR luminosities that
are at least as large as Arp~220 ($>10^{12}~L_\odot$). We therefore
can use the Arp~220 template in making millimetric redshift 
estimates for the optically-faint microjansky sources.

In Fig.~\ref{fig9}a we show \ratio\ versus 1.4\,GHz flux. The
right-hand vertical axis shows the redshift that would be inferred
from the Arp~220 template. In Fig.~\ref{fig9}b
we divide the sample into optically-bright and optically-faint
subsamples. We can see from this figure that the optically-faint 
subsample does indeed move to higher redshift as we probe to 
fainter radio fluxes.


\begin{inlinefigure}
\psfig{figure=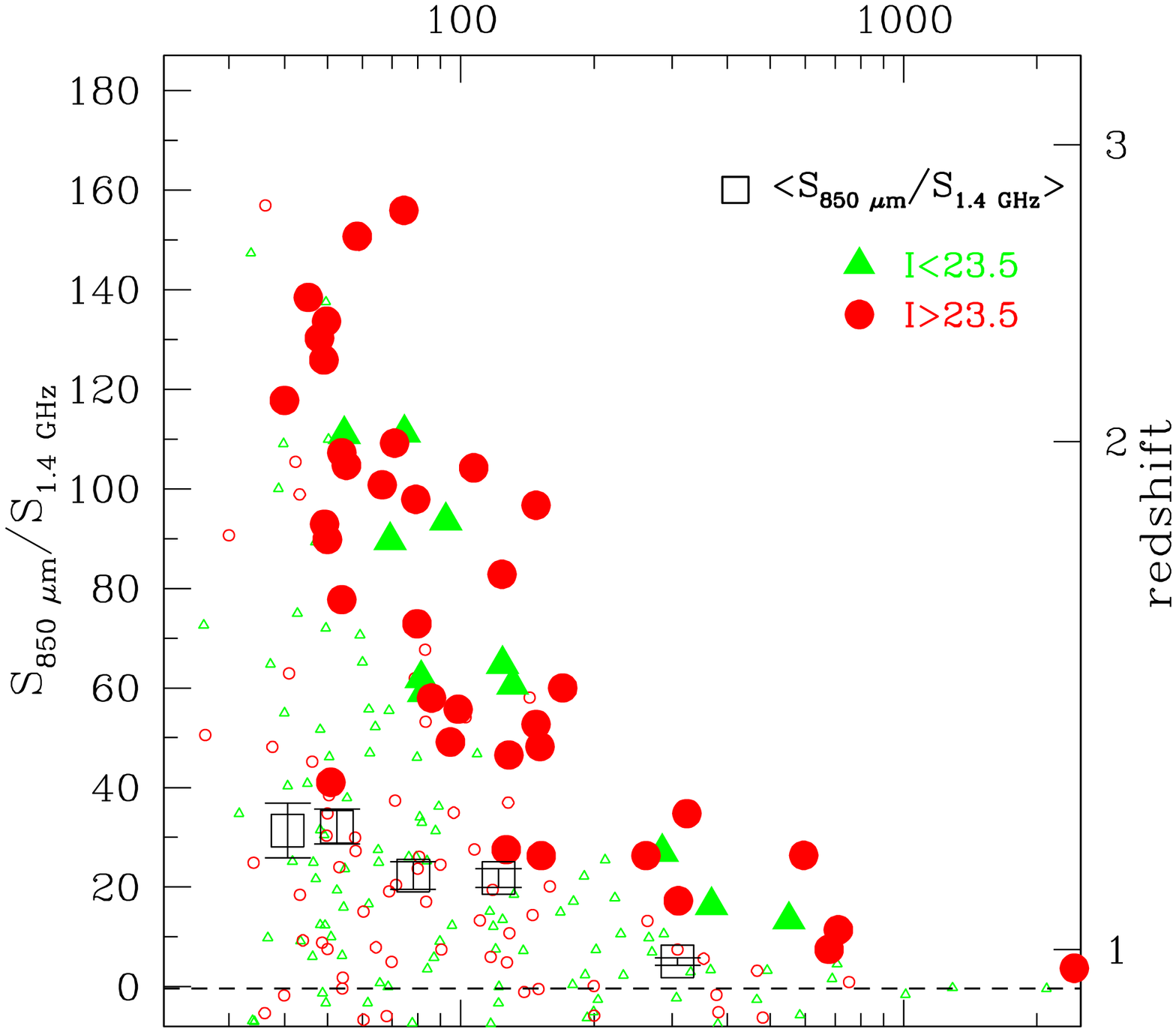,angle=0,width=3.5in}
\psfig{figure=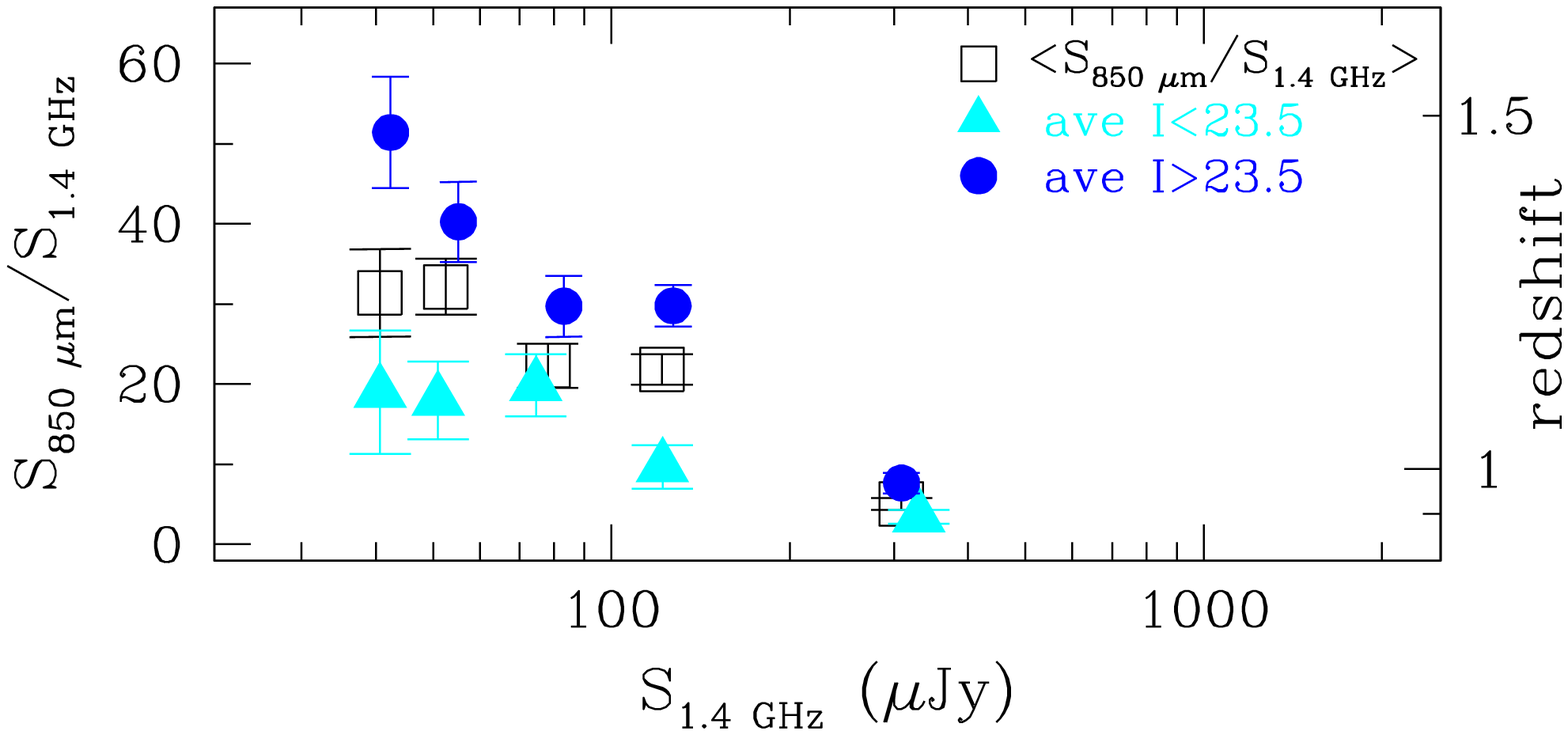,angle=0,width=3.5in}
\vspace{6pt}
\figurenum{9}
\caption{
(a) \ratio\ versus 1.4\,GHz flux for the microjansky
radio sources in the HDF-N, SSA13, and SSA22 fields with
submillimeter measurements. Large filled symbols denote
sources with $>3\sigma$ submillimeter fluxes. Triangles
(circles) denote sources with $I<23.5$ ($I>23.5$).
Assuming the sources span a reasonably small range in
dust temperature and properties, the y-axis can be interpreted
as a redshift measure (right-hand scale). The redshift range covered
by our radio sources is $z\sim0-3$ for typical dust temperatures
encountered in local starburst galaxies and ULIGs.
The mean \ratio\ is independent of radio flux for sources fainter
than about 200~$\mu$Jy, within 1$\sigma$ statistical errors.
(b) Mean \ratio\ versus 1.4\,GHz flux for the optically-bright
and optically-faint subsamples. The nearly
constant \ratio\ ratio of the total sample is dominated by the
optically-bright sources.
}
\label{fig9}
\addtolength{\baselineskip}{10pt}
\end{inlinefigure}

\subsection{Individual sources}
\label{hizids}

There are four significant ($>3\sigma$) submillimeter
sources in our sample with spectroscopic redshifts ($z=3.405$, 
$z=2.565$, $z=1.013$, and $z=0.510$); all have $I<23.5$.
The $z=3.405$ source has Lyman $\alpha$ and CIV in emission, and
the $z=2.565$ source is a broad-line quasar that was detected in
the 100~ks {\it Chandra} exposure of the SSA13 field
(\markcite{barger01a}Barger et al.\ 2001a). The redshifts of 
both sources are very close to their millimetric redshift estimates.
The $z=1.013$ source shows strong Balmer absorption features and
[Ne III] in emission and was detected in the 1~Ms {\it Chandra} Deep
Field-North exposure (\markcite{barger01b}Barger et al.\ 2001b).
The spectroscopic redshifts of this and the $z=0.510$ source are 
lower than their millimetric redshift estimates (assuming an 
Arp~220 SED, we find $z=1.6$ and $z=1.9$, respectively).
Sources with dust temperatures cooler than the Arp~220 $\sim 45$~K 
template will have true redshifts and FIR luminosities lower than 
predicted (see also \markcite{yun02}Yun \& Carilli 2002).
The $z=1.013$ galaxy exhibits a cool dust temperature ($\sim$30\,K), similar to 
recently identified cold ULIGs from the FIRBACK 170$\mu$m ISOPHOT survey
(\markcite{csi02}Chapman et al.\ 2002c).
In the case of the $z=0.510$ galaxy, however, the
dust would have to be prohibitively cold ($\sim18$~K) for the
millimetric redshift to match the spectroscopic identification.
This source may therefore be a misidentification, with the
true source being lensed by the bright foreground galaxy
(\markcite{chapman02b}Chapman et al.\ 2002b).

%
%
\begin{inlinefigure}
\psfig{figure=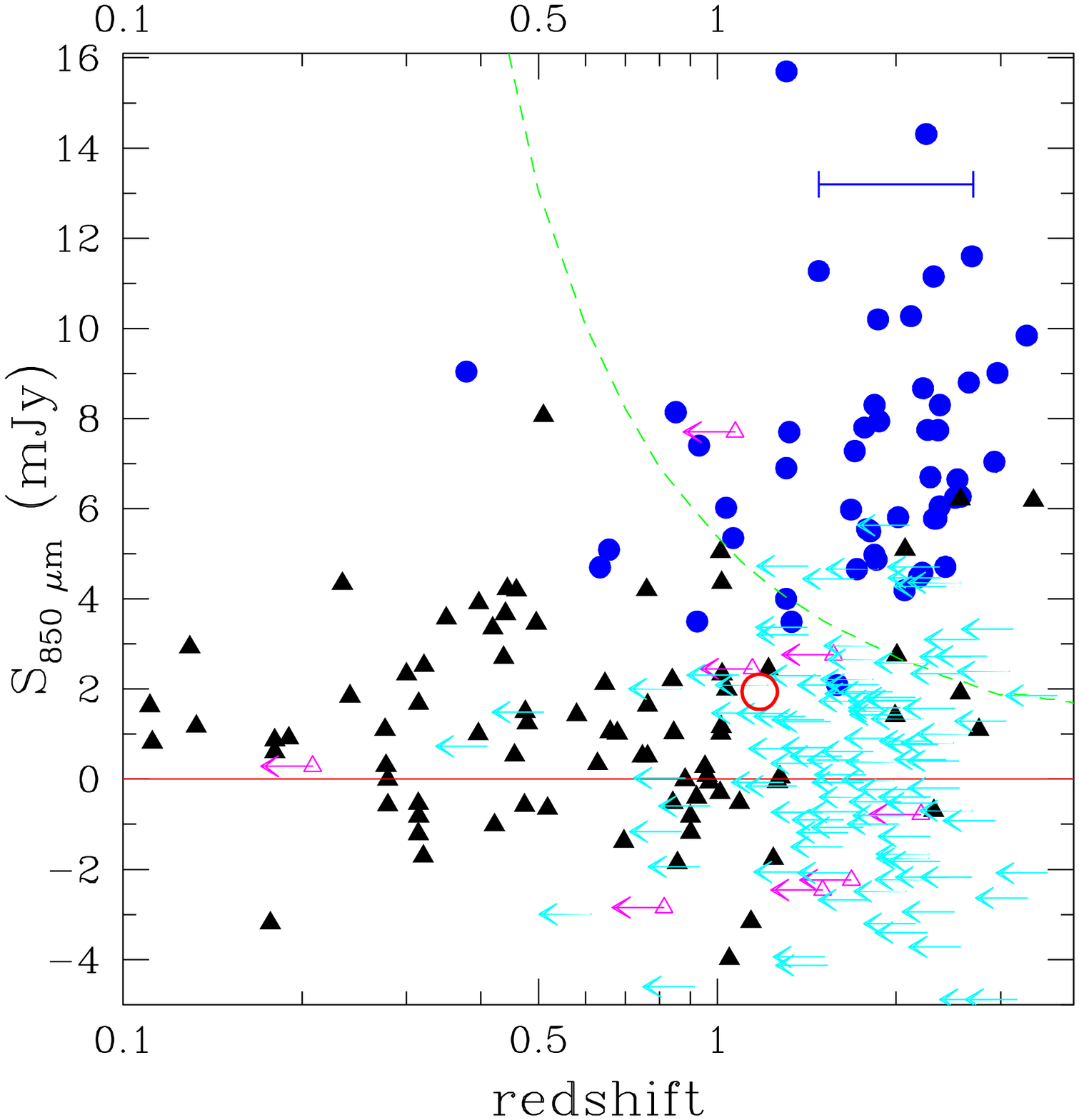,angle=0,width=3.5in}
\vspace{6pt}
\figurenum{10}
\caption{
Submillimeter flux versus redshift for the microjansky radio sources
with submillimeter measurements. The $I<23.5$ sources with
(without) spectroscopic redshifts are denoted by solid (open)
triangles, and the $I>23.5$ sources with \sub\ detections
(and hence millimetric redshifts) are denoted by solid circles.
The sources without $>3\sigma$ \sub\ detections are denoted by
left-pointing arrows at the millimetric redshift
limit (estimated using the $2\sigma$ submillimeter uncertainty)
and the measured \smm\ flux.
The error-weighted mean \ratio\ of these sources is denoted by a
large open circle at its millimetric redshift estimate of $z=1.13$.
The track of a galaxy twice as luminous as Arp~220 is shown as a dashed
line. The expected characteristic error at $z=2$, based on the range in
properties for local ULIGs relative to the Arp~220 template,
is indicated by a horizontal bar.
\label{fig10}
}
\addtolength{\baselineskip}{10pt}
\end{inlinefigure}

\subsection{Combined spectroscopic and millimetric redshift sample}

In Fig.~\ref{fig10} we plot \smm\ versus spectroscopic
(solid triangles) or millimetric (solid circles) redshift 
for the sources with submillimeter measurements.
We also show a rough error bar based on the range in
properties for local ULIGs relative to the Arp~220 template.
We have very little information on the spectroscopically-unidentified
radio sources that are not individually detected in the \sub. We can 
put conservative limits on their redshifts by estimating millimetric
redshifts from the CY00 relation and the $2\sigma$ submillimeter
uncertainty. We denote these sources by the left-pointing arrows
in Fig.~\ref{fig10}.

We can go one step further by determining a single average point
for this marginally detected population
from the ratio of the error-weighted mean \sub\ 
flux and the error-weighted mean radio flux. We denote this
point by a large open circle at $z=1.13$ on Fig.~\ref{fig10}.
As we have already removed all the \smm$>3\sigma$ sources,
submillimeter sources without radio counterparts are not 
present. Since the steep \sub\ source counts imply a large 
Eddington bias, it is important not to overinterpret this result, 
but we note that it falls naturally within the distribution of
the more robustly identified sources.

Comparing the data with the track followed by a ULIG with twice 
the luminosity of Arp 220 (dashed line in Fig.~\ref{fig10}), we can 
see the rapid rise in luminous FIR sources with increasing redshift.
Below $z\sim~1$ there are essentially no such objects,
while above $z\sim~1$ they are common.

\section*{Summary}

We presented optical and submillimeter (jiggle and raster maps
and photometry) imaging and optical 
spectroscopy for a sample of microjansky radio sources detected 
in ultradeep 1.4\,GHz maps of the HDF-N, SSA13, and SSA22 fields.
Our spectroscopic identifications for 169 of the sources in the
HDF-N and SSA13 fields revealed a flat median redshift distribution
across the entire microjansky regime probed. The identified sources 
show a remarkably narrow range in rest-frame absolute $I$-band magnitude
that suggests they are chosen from approximately
$L^\ast$ optical galaxies, regardless of redshift. At the higher redshifts 
the narrow range can be understood from selection effects, but at
the lower redshifts there is no selection against low optical luminosity
galaxies. Thus, the absence of low optical luminosity galaxies 
in the sample argues that they are weak in the radio 
relative to their optical and ultraviolet luminosities, possibly
due to small galaxies not being efficient at retaining
the cosmic rays necessary to host microjansky radio sources.
The fact that these galaxies are eliminated preferentially from a 
flux-limited radio survey poses a serious problem for radio estimates 
of the local star formation rate density since a substantial fraction 
of the ultraviolet luminosity density is generated by sub-$L^\ast$ 
galaxies at low redshifts.

From a statistical analysis of our submillimeter measurements for 278 
of the microjansky radio sources in the HDF-N, SSA13, and SSA22 fields, 
we found that optically-faint radio sources are about three times more
submillimeter luminous, on average, than optically-bright radio
sources, but both populations are, on average, very significantly 
detected in the submillimeter. Fifty of the radio sources
in our sample are significantly ($>3\sigma$) detected individually 
in the submillimeter; 38 of these are optically faint ($I>23.5$).
We have spectroscopic redshifts for four of the 
$I<23.5$ submillimeter-detected sources, three of which
are in the range $z=1$--3.4; the fourth source is at
$z=0.510$ and may be a misidentification.

To quantify the percentage of optically-bright and 
optically-faint radio sources that are significantly detected in
the submillimeter, we considered only the regions of our survey
that were mapped with SCUBA (as opposed to the regions that were
targeted with photometry). We found that $69\pm9$\%
of the radio-submillimeter population has $I>23.5$. 
We also found that $66\pm7$\% of the \smm$>5$\,mJy 
($>4\sigma$) sources in our fields are radio-identified.

We used our spectroscopically-identified microjansky radio sample 
to determine the evolution with
redshift in the radio power and hence the FIR luminosity 
(through the FIR-radio correlation). We found that
ULIGs or near-ULIGs are extremely rare at low redshifts
but are relatively common by $z=1$. In combination with the
lower-bound selection effects, this means that we have a population
which is rapidly changing with redshift but that is fairly
tightly selected at each redshift.
At low redshifts ($z<0.4$) we are selecting Milky Way type galaxies,
at intermediate redshifts we are selecting LIGs, and at high
redshifts ($z>1$) we are selecting only ULIGs or near-ULIGs.
From the error-weighted mean \ratio\ flux ratio we see the same 
evolution with redshift, such that at lower redshifts the 
error-weighted means follow the track of a quiescent to
moderately star-forming galaxy, while at higher redshifts the
observations migrate to values consistent with Arp~220.
Thus, millimetric redshift estimates at low redshifts are best
made with a template intermediate between a Milky Way type
template and a starburst template, while at high redshifts 
($z>1$) an Arp~220 template is appropriate.
With this knowledge, we estimated millimetric redshifts or
limits for all of the optically-faint microjansky radio
sources in our sample with submillimeter measurements.

\acknowledgements
We thank an anonymous referee for helpful comments that improved 
the manuscript. We thank Ian Smail for a careful reading of a 
draft of the paper and for useful discussions.
We gratefully acknowledge support from NASA through
grant 9174 (SCC) and DF1-2001X (LLC), from the
University of Wisconsin Research Committee with funds
granted by the Wisconsin Alumni Research Foundation (AJB),
and from NSF grants AST-0084847 (AJB) and AST-0084816 (LLC).
DS and CB are supported by the Natural Sciences and Engineering
Research Council of Canada.

\newpage

\begin{deluxetable}{lcccc}
\renewcommand\baselinestretch{1.0}
\tablewidth{0pt}
\parskip=0.2cm
\tablenum{1}
\tablecaption{Average Submillimeter Fluxes for the HDF-N+SSA13+SSA22 Sample
in $I$-magnitude bins}
\small
\tablehead{
\colhead{$I$-mag Bin}
&  \smm\ (mJy)
& \colhead{$\sigma_{850}$ (mJy)}
& number in bin
& \% SMM detections ($>3\sigma$) \cr
}
\startdata
$I>25$     &  2.38  &    0.18  &           68 & 26\%\\ 
$24<I<25$  &  2.85  &    0.23  &           43 & 37\%\\ 
$23<I<24$  &  1.32  &    0.27  &           32 & 24\%\\ 
$22<I<23$  &  1.00  &    0.28  &           36 & 8\%\\ 
$21<I<22$  &  1.13  &    0.26  &           31 & 9\%\\ 
$19.5<I<21$&  1.10  &    0.31  &           31 & 3\%\\ 
$I<19.5$   &  1.08  &    0.29  &           37 & 5\%\\ 
\enddata
\label{tab1}
\end{deluxetable}

\end{document}